\documentclass[aps,prb,twocolumn,eqsecnum,showpacs]{revtex4}

\usepackage{amsmath}
\usepackage{amssymb}
\usepackage{amsfonts}
\usepackage{bm}
\usepackage{graphicx}
\usepackage{subfigure}

\begin{document}

\title{Langevin Dynamics of the vortex matter two-stage melting transition in
  Bi$_2$Sr$_2$CaCu$_2$O$_{8+\delta}$ in the presence of straight and
  of tilted columnar defects}

\author{Yadin Y. Goldschmidt and Jin-Tao Liu}

\affiliation{Department of Physics and Astronomy, University of Pittsburgh,
Pittsburgh, Pennsylvania 15260}

\date{\today}

\begin{abstract}
In this paper we use London Langevin molecular
dynamics simulations to investigate the vortex matter melting transition
in the highly anisotropic high-temperature superconductor material
Bi$_2$Sr$_2$CaCu$_2$O$_{8+\delta}$ in the presence of low
concentration of columnar defects (CDs). We reproduce with
further details our previous results obtained by using Multilevel
Monte Carlo simulations that showed that the melting of the
nanocrystalline vortex matter occurs in two stages: a first stage melting into
nanoliquid vortex matter and a second stage delocalization transition into a
homogeneous liquid. Furthermore, we report on new dynamical measurements 
in the presence of a current that identifies clearly the
irreversibility line and the second stage
delocalization transition. In addition to CDs aligned along the c-axis we also
simulate the case of tilted CDs which are aligned at an angle with respect to
the applied magnetic field. Results for CDs tilted by $45^{\circ}$ with
respect to c-axis show that the locations of the melting and
delocalization transitions are not affected by the
tilt when the ratio of flux lines to CDs remains constant. On the
other hand we argue that some dynamical properties and in particular 
the position of the irreversibility line should be affected.
\end{abstract}
\maketitle

\section{Introduction}

Since their discovery during the 1980s, much progress has been made on the study of
high-temperature superconductors. These materials are classified as type II
superconductors, for which there is only a partial Meissner effect
when the external magnetic 
field is in the range $H_{c1}<H<H_{c2}$ \cite{tinkham,blatter,brandt}. 
Inside the superconductors, the magnetic field survives in the form of
quantized flux lines (FLs), called vortices,
each with a quantum unit of flux $\phi_0 = hc/2e$.
The vortices form a periodic hexagonal lattice (Abrikosov lattice) at low temperatures
and melt into a vortex liquid at higher temperatures
\cite{exp1,exp2,cubitt,exp3,exp4,crabtree}. 

High-temperature superconductors are ceramic materials which have the common
structure of layered superconducting copper-oxide planes. They are anisotropic materials,
with anisotropy defined as $\gamma = \sqrt{m_z/m_{\bot}}$, where $m_z$ and $m_{\bot}$
are the effective masses of electrons moving perpendicular to the superconducting
planes and along to them respectively. One of the commonly studied materials
Bi$_2$Sr$_2$CaCu$_2$O$_{8+\delta}$, known as BSCCO, is highly anisotropic. It is estimated
to has an anisotropy in the range of 300-500 for optimally doped samples.

In high anisotropic materials, vortices can be described as stacks of pancake
vortices \cite{clem,clemnew}. The ``pancakes'' are centered at their
corresponding planes and 
can only move within their own planes. The interaction between the pancakes consists of
two parts: the electromagnetic interaction and the Josephson interaction.

The electromagnetic interaction originates from the screening currents that arise in the
same plane where a pancake resides as well as in more distant planes. It is
repulsive among pancakes in the same plane and attractive for those in different
planes \cite{artemenko,clem}. This interaction exists even in the case
where the superconducting 
planes are completely decoupled, so no current can flow along the $c$ axis (perpendicular
to the planes) of the sample .

The Josephson interaction \cite{artemenko,clem1,blatter} originates from the Josephson current flowing between
the superconducting planes, which is proportional to the sine of the gauge invariant
phase difference between two planes $\Delta \varphi + (2\pi/\phi_0) \int d{\bf s \cdot A}$,
where $\varphi$ is the phase of the superconducting order parameter and ${\bf A}$ is the
vector potential. When two pancakes belonging to the same stack and residing in adjacent
planes move away from each other, the phase difference that originates
causes a Josephson current to begin flowing between the planes. This
results in an attractive interaction between pancakes that for
distances small compared to $\Lambda \equiv \gamma d$ is approximately
quadratic \cite{artemenko,blatter} in the distance. Here we denoted by
$d$ the inter-plane separation. When the two adjacent pancakes are
separated by a distance larger than 
$\Lambda$, a ``Josephson string'' is formed, whose energy is proportional 
to its length \cite{clem1,koshelev}.

Recently the effect of columnar defects (CDs) on the melting transition
was investigated both experimentally
\cite{banerjee,menghini,banerjee2} and  numerically 
\cite{styyg1,dasgupta,nonomura,goldcuan}. 
Columnar defects can be induced artificially by irradiating
the sample with high energy heavy ions, like 1 GeV lead ions. The heat
released by the ions create tubes of damaged, non-conducting material.
It is customary to measure the density of CDs by the value of the
``matching field'' by $B_\phi=n_{cd} \phi_0$, where $n_{cd}$ is 
the number of CDs per unit area. In the limit $B \ll B_\phi$ the
vortex solid becomes a Bose glass as most FLs are trapped by CDs
located at random positions \cite{blatter,Nelson,Radzihovsky,Kees}. 
On the other hand our interest is in
the case $B_\phi\le B$, in which there are more FLs than CDs. In this
case the picture that
emerged from experiments \cite{banerjee,menghini,banerjee2} and from 
numerical investigations \cite{styyg1,dasgupta,nonomura,goldcuan} and analytical
calculations \cite{lopatin,goldcuan} is that
of the ``crystallites in the pores'', also referred to as the ``porous''
vortex solid. At low temperatures a skeleton (or matrix)
of vortices localized on CDs is formed, while the excess (interstitial) FLs
try to form hexagonal crystallites in the lacuna between CDs. Thus this phase
has a short ranged translational order, which extends to a distance of
the order of a typical pore size. 

As the temperature is increased, and
if the magnetic field is large enough, this heterogeneous structure
melts in two stages: first the crystallites in the pores melt into a
nanoliquid while the skeleton remain intact, and subsequently the
skeleton melts and the liquid becomes homogeneous. When the magnetic
field is lowered these two
transitions coincide into one. This is usually associated with a kink
in the first order melting line. The first stage melting of the
crystallites is observed in the experiments as a step in the
equilibrium local magnetization \cite{Pastoriza,Zeldov,Schilling} 
and in the simulations as a
sharp increase in the transverse fluctuations of the FLs, among other
signatures. The second transition was not observed in the
experiments as a similar jump in the magnetization or another
equilibrium property. It was observed in transport measurement \cite{banerjee2}
involving transport currents with alternating polarity. This
establishes the transition as dynamical in nature. It remains to be
established that this is also an equilibrium thermodynamic
transition. In order to support this assertion  it was argued \cite{banerjee2}
that the second ``delocalization'' transition is associated with the
restoration of a broken longitudinal gauge symmetry. However, at this
time it cannot be ruled out that the second transition is only a crossover
associated with a gradual equilibrium change over a finite range of
temperatures. 

Recently one of us published a short paper \cite{goldcuan} reporting
on multilevel Monte Carlo simulations of the vortex matter in the 
highly anisotropic high-temperature superconductor BSCCO. We
introduced low concentration of columnar defects
satisfying $B_\phi \le B$. Both the electromagnetic and Josephson interactions
among pancake vortices were included. The nanocrystalline, nanoliquid,
and homogeneous liquid phases were identified in agreement 
with experiments. We observed the two-step melting process and also
noted an enhancement of the structure factor just prior to the melting
transition. We also proposed a simple theoretical model to explain
some of the findings. In this paper we would like to give more details
of the simulation results but we use a different method than in the
previous work, i.e. we use molecular dynamics instead of multilevel
Monte carlo. This enables us to carry out dynamical measurements that
clearly identifies the second melting transition (delocalization). We
also report in this paper simulations of tilted columnar defects. A
short summary of our findings for tilted defects has recently been
submitted \cite{nurit}. 

Molecular dynamics (MD) is a powerful tool for simulations of physical
systems and it often serves as an alternative to Monte Carlo (MC)
simulations. Its advantage is that it can be used to investigate the
real dynamics of the system as opposed to MC simulations that are used
for obtaining equilibrium properties. However MD simulations could be
plagued by the absence of ergodicity when applied to systems
represented by path integrals \cite{ceperley} and there is also the
problem of implementing permutations for the case of identical
particles like Bosons. The problem of ergodicity is really not much of
an issue for Langevin simulations since the thermal noise helps the system
explore the configuration space and it can be shown by using the
corresponding Fokker-Planck equation that equilibrium is reached in
the long time limit. For flux-lines we have found a way to
implement ``permutations'' in the MD simulations by flux cutting and 
recombining and this was explained in a recent publication
\cite{yyg} which reported on MD simulations of the melting transition
in pure BSCCO without columnar or point defects. We were also able to 
implement periodic boundary conditions in all directions 
(including the $z$ direction), and to include the in and out of 
plane electromagnetic interaction as well as the Josephson interaction
using the new approximation we have recently obtained \cite{yygst3}.
Results for the first-order melting transitions in
BSCCO for fields of 100-200 gauss were found to be in
an excellent agreement with the results of our previous multilevel Monte Carlo
simulations \cite{styyg2} including the proliferation of non-simple  
loops corresponding to flux entanglement above the melting transition.
In this paper we extend the MD simulations to systems with columnar
defects, both for the case that the columns are parallel to the c-axis
and for the case when the columns are tilted at an angle $\theta$ with
respect to the c-axis. The external field is taken to be aligned along
the c-axis. We would like to check if the location of the melting
transition in the B-T plane is sensitive to the tilting angle of the
columnar defects. Recently experiments were carried out to investigate
this problem \cite{nurit}. The conclusions were that while static
thermodynamical signatures like the location of the melting and
delocalization lines do not depend (within the experimental errors)
on the angle between the applied field and the columnar defects, 
non-equilibrium properties like the position of the irreversibility
line do depend on the angle.

\section{ Simulation method}

\subsection{method}

In this paper we use the London-Langevin molecular dynamics simulation
method. The details of the simulation method, for a system without
columnar defects is given in our previous paper \cite{yyg}. Here we
summarize the main ingredients of the method. 
The equation of motion for the $m$'th pancake vortex is
\begin{eqnarray}
d\ \eta\ \frac{d{\bm R}_m}{dt}=-{\bm\nabla}_m V(\{{\bm
  r}_n\})+{\bm f}_L+{\bm \zeta}_m(t).
\label{langevin}  
\end{eqnarray}
The pancake label $m$ stands actually for two indices $(i,p)$ where
$p$ is the plane label and $i$ is the pancake label in that plane.
The position ${\bm R}_m$ is a two component vector in the plane.
We use the over-damped model for vortex motion in which the
velocity of the vortex is proportional to the applied force and $\eta$
is the viscous drag coefficient per unit length given by the
Bardeen-Stephen expression \cite{bardeen}.
In Eq.(\ref{langevin}) $d$ is the interlayer
spacing between CuO$_2$ planes that is taken to be equal to the width
of the pancake vortex. $V$ is the potential
energy depending on the position of all pancakes and includes both the
magnetic energy and Josephson energy, and the interaction between a
pancake and the columnar defects. The force is minus the
gradient of the potential energy with respect to the position of the
$m$'th pancake. ${\bm f}_L$ is a driving force (if present), for example
the Lorentz force induced by a current. ${\bm \zeta}_m$ is a white thermal
noise term which satisfies
\begin{eqnarray}
  \langle \zeta^\alpha_m(t)\zeta^\beta_n(t')\rangle=2kT\eta d\
  \delta_{\alpha \beta}\delta_{m n}\delta(t-t').
\label{noise}
\end{eqnarray}
In Eq.(\ref{noise}) $\alpha$ and $\beta$ refer to the $x$ and $y$
components of the vector ${\bm \zeta}$ and $m$ and $n$ are pancake
labels. $k$ is Boltzmann's constant. In the simulations we use a
discrete form of the Dirac delta-function.
We measure distances in units of
$a_0=\sqrt{2 \phi_0/B \sqrt{3}}$ where $B$ is the magnetic field. We
measure energy in units of $\epsilon_0 d$
where $\epsilon_0(T)=(\phi_0/4\pi\lambda)^2$ is the basic energy scale per
unit length and $\lambda$ is the penetration depth. We measure time in
units of $\eta a_0^2 /\epsilon_0$. For example for $T=60K$ and
$B=100G$, $a_0\approx 4887$ \AA,
$\epsilon_0 d\approx 4.685\times 10^{-14}$erg $\approx 339.5$
K/k. Ref.~[\onlinecite{bulaevskii}] quotes a value for $\eta$ for a
single crystal BSCCO of around $1\times 10^{-7}$ g/(cm s). Based on
this value the time unit is about $0.765$ ns. These values change when
the temperature and field are varied.
The simulation cell was chosen to have a rectangular cross section of
size $L_1\times L_2=a_0\sqrt{N_{fl}}\times a_0\sqrt{3N_{fl}}/2$ where $N_{fl}$ is
the number of flux lines (number of pancake vortices in each
plane). We usually worked with 36 flux lines but to check for finite
size effect we also used 64 FLs in some simulations. The aspect ratio of the
cell was chosen to accommodate a triangular lattice without distortion,
such that each triangle is equilateral. In the z-direction we take
$N_{p}$ layers of width $d$ each, where in practice we have chosen
$N_{p}$ to be 36 or 200, as will be indicated below. Thus the
number of pancakes used is from 1296 up to 7200. Note that our system
is effectively bigger since every pancake interacts not only with all
other pancakes in the simulation cell but with infinitely many
images of these pancakes in neighboring cells in all directions.

\subsection{interactions}

The electromagnetic (EM) interaction occurs between any pair of
pancakes. To produce more realistic results for the small system
simulated we have implemented periodic boundary conditions (PBC) in all
directions including the z-direction.  Periodic boundary conditions
mean that every pancake interact not only with the actual pancakes in
the simulation cell but also with all their images in other cells which are
part of an infinite periodic array. Each image of a pancake is located
at the same position in the corresponding cell as the original pancake
in the simulation cell. The details are explained in Ref.~[\onlinecite{yyg}]. 
The final expression we used for the pair energy of two pancakes
with fully implemented PBC is
\begin{eqnarray}
\label{eq:pbcout}
\frac{U_{\!\rm mag}(\!{\bm R},\!\Delta p\!\!\neq\!\! 0)}{\epsilon_0d}\!\approx\!
      \frac{d}{\lambda}f_m\!(\Delta p)\!\!
\left[\!G_{\!0}\!\!\left(\!\frac{\bm R}{\lambda},\!\frac{L_1\!}{\lambda}\!\right)
\!\!-\!G_{\!0C}\!\!\left(\!\frac{x}{L_1\!},\!\frac{y}{L_2\!}\right)\!\right]\!,\nonumber\\
\end{eqnarray}
for pancakes in different planes separated by a vertical distance
$d\Delta p$ and a lateral separation $\bm R=(x,y)$, and similarly
\begin{eqnarray}
  \label{eq:pbin}
\frac{U_{\!\rm mag}({\bm R},0)}{\epsilon_0d}
\!\!&\approx&\!\!
  2G_{0C}\left(\frac{x}{L_1},\frac{y}{L_2}\right) \nonumber \\
& & \!\!+\frac{d}{\lambda}f_m\!(0)\!
  \left[\!G_{\!0}\!\left(\!\frac{\bm R}{\lambda},\frac{L_1\!}{\lambda}\right)
  \!\!-\! G_{\!0C}\!\left(\frac{x}{L_1\!},\frac{y}{L_2\!}\right) \!\right]\!,\nonumber\\
\end{eqnarray}
for pancakes in the same plane. The functions $G_0({\bm
R/\lambda},{L_1}/{\lambda})$ and 
 $G_{0C}({x}/{L_1},{y}/{L_2})$ represent the periodic screened and
 bare Greens functions for the Coulomb interactions in 2D and are
 given in the Appendix of 
 Ref.~[\onlinecite{yyg}]. The factor $f_m$ comes from the summation of
 the EM interaction between a pancake and another pancake in the same cell and all its
 images along the z-direction. It is given by
\begin{eqnarray}
f_m(\Delta p)
\!\!&=&\!\! \sum_{l=-\infty}^\infty \exp(-|\Delta p+N_p l|\mu)\nonumber \\
\!\!&=&\!\! \frac{\exp(-|\Delta p|\mu)+\exp((|\Delta p|\!-\!N_p)\mu)}{1-\exp(-N_p\mu)} 
\label{fm}
\end{eqnarray}
where we put $\mu=d/\lambda$. The dependence of
$f_m$ on $\Delta p$ is rather weak.
The approximations are valid provided $dN_p<\lambda$.

We have also implemented the Josephson interaction among adjacent
pancakes belonging to the same FL \cite{yyg}.
The approximate formula we use is based on our recent paper that
derives a numerical solution to the nonlinear sine Gordon equation in
two dimensions \cite{yygst3}. The formula interpolates between an
$\epsilon_0d (R/\Lambda)^2\ln(\Lambda/R)$ dependence for small $R$ to
 $\epsilon_0d R/\Lambda$ dependence for large $R$, where $R$ is the
lateral separation of the pancakes, $\Lambda=\gamma d $, $\gamma
$ is the anisotropy and $d$ is the inter-plane separation. 
This interpolation is an modification of a similar interpolation
formula previously used by Ryu, Doniach, Deutscher and Kapitulnik
\cite{ryu}. The Josephson interaction diminishes with increasing
anisotropy, and we choose the anisotropy parameter to be 375 to obtain
reasonable agreement with recent experiment on pristine melting. In
this work we neglect Josephson interaction between pancakes belonging
to different stacks because on the average those pancakes are farther
away. In addition we view the stack as the collection of pancakes
carrying the unit quantum of flux from one side of the sample to the
other. Since the josephson interaction is 
proportional to the gauge invariant phase difference $\Delta \varphi +
(2\pi/\phi_0) \int d{\bf s \cdot A}$, the second term involving the
vector potential will be larger for pancakes belonging to the same
stack than to pancakes belonging to different stacks. This argument 
works in the solid phase but not deep into the liquid phase where eventually 
stacks may lose their meaning, especially if the density of pancakes is large 
so the average distance is much smaller than $\lambda$ (high temperatures and high fields).     

Since the Josephson interaction is between nearest neighbor pancakes in
adjacent planes it is quite straight forward to implement PBC. A pancake at
the top plane ($N_p$) interacts with the closest pancake in the bottom
plane $1$ as well as 
with a pancake in plane $N_p-1$. When calculating the lateral distance
between pancakes we always measure the ``shortest distance'' defined as
follows: If the actual $|\Delta x|$ separation is larger than $L_1/2$ we
subtract or add $L_1$ depending on the sign of $\Delta x$, and
similarly for $\Delta y$ with $L_2$ replacing $L_1$. This way the
correct distance is obtained even when the adjacent pancake in the
plane above has exited the simulation cell and emerged
close to the other side of the simulation cell.

In the case of $R \gg r_g$, string-string interactions that involve
three and four-body interactions become important \cite{bulaevskii2}. However near
the melting transition for the range of magnetic fields investigated in this
paper $R \approx 0.25 a_0 \approx 1000$\AA\ whereas $ r_g\approx
5625$\AA. Thus large transverse 
fluctuations for which the string-string interactions become
important are statistically rare and can be neglected. Even in the
case of tilted CDs we verified that the kinks present (see later
discussion) are not fully developed Josephson strings. Thus in this
work we do not include string-string interactions.

The modeling of
the interaction between pancakes and CDs was discussed by
Goldschmidt and Cuansing \cite{goldcuan}. There are two major sources
of pinning: core pinning and EM 
pinning. Core pinning \cite{blatter} arises when the vortex core overlaps
with a normal state inclusion similar to the one inside a CD. Since
condensation energy is lost in the vortex core, part or all of this
energy is restored when a vortex resides inside a
CD. EM pinning arises \cite{Buzdin,Mkrtchyan} when the
supercurrent pattern around the vortex is disturbed by the
non-conducting defect. These two mechanisms combine
together to yield the expression for the potential energy at a
distance $R$ away from the CD as felt by an individual pancake
\cite{blatter,Mkrtchyan}: 
\begin{eqnarray}
V(R)\approx \left\{ \begin{array}{c}\epsilon_0 d\
\ln\left[1-\left(\frac{r_r}{R+\xi/\sqrt{2}}\right)^2\right],\ \ R>r_r,
\\
\epsilon_0 d\
\ln\left[1-\left(\frac{r_r}{r_r+\xi/\sqrt{2}}\right)^2\right],\ \ R<r_r,
\end{array}\right.   \label{eq:bind}
\end{eqnarray}
where $r_r$ is the radius of the CD. The long range tail 
contribution $\approx
-\epsilon_0 d r_r^2/R^2$ is due mainly to the EM pinning
and the short range flat region of depth $\approx \epsilon_0 d$ is due
to the combined effects of EM and core pinning. This 
formula is valid provided $r_r>\sqrt{2}\xi$. If $r_r<\sqrt{2}\xi$ a slightly
different expression should be used, but we have determined, see below,
that in order to get good agreement with experiments we need to use
$r_r>\sqrt{2}\xi$. Note that in this expression $\epsilon_0$ is
temperature dependent and is proportional to $1-T/T_c$. For example the pinning strength
reduces to realistic values of 113 K at T=80 K and 79K at T=83K
assuming $T_c=90$K.

After experimenting with CDs of different radii ($10$ nm $\le r_r\le
30$ nm) we concluded that in 
order to obtain a good agreement with experimental results we needed to
choose $r_r \approx 30$ nm in which case the formula given by
Eq.~(\ref{eq:bind}) is the 
correct one. Different heavy ions with different energies give rise
to tracks of various sizes, normally reported to be in the range of
$4-20$ nm in diameter. It may be that the actual damage caused by a track
exceeds its apparent size, thus leading to a higher effective radius.
For tilted CDs it may be appropriate to take a potential with an
elliptical cross section rather than a circular one. Since our simulations
for tilted CDs are mainly for angles of $45^{\circ}$ 
which are not that large and since the radius of our
CDs is already larger than the actual experimental radius, we did not
use elliptical cross sections in the present simulations.

\subsection{Details of the Simulations}

The simulation cell is divided into a $800\times692$ mesh of small cells 
of area $\tilde{h}\times \tilde{h}$ each, where 
$\tilde{h}=\sqrt{N_{fl}}/800$ (in units of $a_0$). We
tabulate the functions $G_0$ and $G_{0C}$ using a less refined cell structure
creating two large $200\times 174$ matrices. During the simulations we use the tables
as a lookup to calculate the pair interaction and use interpolation
to find the applicable values. For each table we also
calculate the negative of the gradient and save the two components of
the gradient in their own tables. We also tabulate the Josephson interaction
and its gradient and the columnar defect potential and its gradient
but for these quantities we used a $800\times692$ matrices. When simulating
we allow pancakes to move to
arbitrary real locations, but in order to calculate the forces we
divide the actual position by $\tilde{h}$ and round to the nearest integer to use for
the lookup tables.

In each simulation step we move all the pancakes at the same time,
using the instantaneous forces. This is done using a time step $\Delta
\tilde{t}$. 
In these simulations we used the second order Runge Kutta
method to advance the solution of the differential equations, which
requires an additional virtual move for any real move. 
It is very important to chose the time step correctly, especially when
CDs are present. We used a variable size time step that limit how much a
pancake can move in each step. The maximum distance allowed was 0.06
$a_0$. Thus if the maximum force on the
pancakes is very large, a small time step is used.

In the simulation we apply a procedure to implement flux cutting and
recombination. We assume that within the coupled-planes model,
vortices may switch connections to lower their elastic energy (in this
case Josephson energy) when they cross each other \cite{ryu}.
In the simulation we construct two matrices of size
$N_{fl}\times N_p$ which we call the ``up'' matrix and the ``down''
matrix. For a 
given pancake $i$ in plane $p$ the ``up'' matrix points to the pancake in
the plane $p+1$ (or 1 if $p=N_p$) that is connected to the given
pancake $(i,p)$ via a Josephson interaction. Generally this is the
pancake closest to the given pancake in the next plane. The ``down''
matrix similarly points to the closest pancake below (or in plane
$N_p$ for $p=1$). When we start from an initial configuration in which
the FLs are straight stacks of pancakes, the matrices simply point to
the pancake just above or below a given pancake. 
When constructing the force matrix after each time
step we check if indeed the ``up'' matrix points to the closest
pancake above. If there
is a closer pancake than the one given by the pointer then we find out its
parent in the plane $p$ by using the down matrix, and we check if 
switching the two connections will decrease the sum of the squares of
the two distances. If it does we cut and switch connections and update
the ``up'' and ``down'' matrices. We term this process an ``{\it
  exchange}''. The reason we use the square of the
distances is that in most instances the Josephson interaction is
proportional to the square of the transverse distance.
This procedure mimics the actual dynamics in which we expect
the magnetic flux to choose a path that minimizes the
Josephson energy. We implement the flux cutting procedure after every
update move of the system, but not during the virtual half-step in the
Runge-Kutta procedure.

Our simulations show
that just below the melting transition, even though some exchanges occur,
they soon reverse or undo themselves, and thus they do not
lead to what we refer to as an entangled state where composite loops
or permutations are abundant. On the other hand
when exchanges proliferate through the system, a phenomenon that occurs
in our simulations just above the melting transition, the exchanges do
not undo each other, and the system of FLs
changes from being composed of simple loops each made up of a single
FL ending on itself,
to a system composed mainly of composite loops that wind up several
times around the simulation cell in the $z$ direction before returning
to the original point. The reason that the exchanges proliferate above
the melting transition is that the transverse fluctuations become
strong enough to overcome the potential barriers due to the repulsion
among pancakes residing in the same plane and thus the crossing of
FLs occur.

\subsection{Measured quantities}
We measured the following physical quantities. For details the reader
is referred to our earlier work \cite{styyg1,styyg2}. The
translational structure factor was also measured but we do not report
on it in the present paper. 

\subsubsection{Energy}

We calculated the average energy per pancake. The total energy is the sum of the
contributions from the EM energy of all pairs of pancakes, the
Josephson energy of nearest neighbor pancakes in adjacent planes and the binding
energy of pancakes trapped by columnar defects. For the first part, the EM energy
of a perfect Abrikosov lattice at the same temperature and magnetic field was
subtracted from the total EM energy \cite{nordborg}.

\subsubsection{Mean square deviations}

For each individual flux-line we define the position of the
lateral center of mass as ${\bm R}_{CM}=\sum {\bm R}_{(i,p)}/N_p$,
where the sum goes over all the pancakes belonging to it. We then define
the mean square deviations as
\begin{eqnarray}
  \label{eq:rmsf}
  R_{f}^2=\left\langle\sum({\bm R}_{(i,p)}-{\bm
        R}_{CM})^2\right\rangle, 
\end{eqnarray}
Where the sum is over all pancakes belonging to an individual
flux-line and the average is over all flux-lines of the system and
then taking a time average. The
melting transition is expected to occur when this quantity satisfies
\begin{eqnarray}
  R_f/a_0 \ge c_L,
\label{lindemann}
\end{eqnarray}
where $c_L$ is the Lindemann coefficient.

\subsubsection{Line entanglement} 

As we allow for flux cutting and recombination, we can define the
number $N_{e}/N_{fl}$ 
as that fraction of the total number of FLs which belong to loops that
are bigger than the size of a ``simple'' loop. A simple loop is
defined as a set of $N_p$ beads connected end to end (due to the
periodic boundary conditions in the $z$ direction).
Loops of size $2N_p$, $3N_p$... start proliferating at and above the melting
temperature.

\subsubsection{drift velocity under an applied force}  
For the dynamical measurements, a uniform driving force of proper magnitude was
applied along positive $x$ direction. The average drift velocity of each pancake
was measured. Both the distribution and the average value of these velocities are
studied. They provide evidence on the delocalization transition and the melting
transition.

\subsubsection{Parameters}

Parameters for BSCCO were taken as follows:
$\lambda_0=1700${\AA}, $d=15${\AA}, $\xi_0=30${\AA} and $T_c=90$ K. 
The temperature dependence of $\lambda$ in this work was taken to follow the
Ginzburg-Landau convention $\lambda^2(T)=\lambda_0^2/(1-T/T_c)$ and
similarly for $\xi$. See discussion in Ref.~[\onlinecite{styyg2}] on
the agreement of this choice
with experiments. For the anisotropy $\gamma$ we have used value of
375. For columnar defects we have used a radius of 30nm.

\section{vertical columnar defects}

In this section, we concentrate on the simulations of systems with vertical CDs.
Both static measurements and dynamical measurements have been done.
For the static measurements, we start from a hexagonal lattice of straight and vertical
FLs. The CDs attract those FLs which are near them and rapidly become occupied.
For temperatures close to and above the melting transition, the FLs also become entangled.
It takes relatively long for the entangling process to reach equilibrium, especially near the transition.
We took 50 units of model time (time is measured in unit of $\eta a_0^2 \epsilon_0$)
for equilibration and another 50 units of model time on measurements.
Simulations have been done on systems with 36 FLs and 36 layers.
Columnar defect density $B_{\phi}$ was set to be 50G.
Magnetic fields from 50G to 200G were investigated. For each field value,
10 different CD configurations were simulated and the measurement results were averaged.

\begin{figure}
\subfigure[]{
    \label{fig01_a}
    \begin{minipage}[b]{0.5\textwidth}
    \centering
    \includegraphics[width=2.8in]{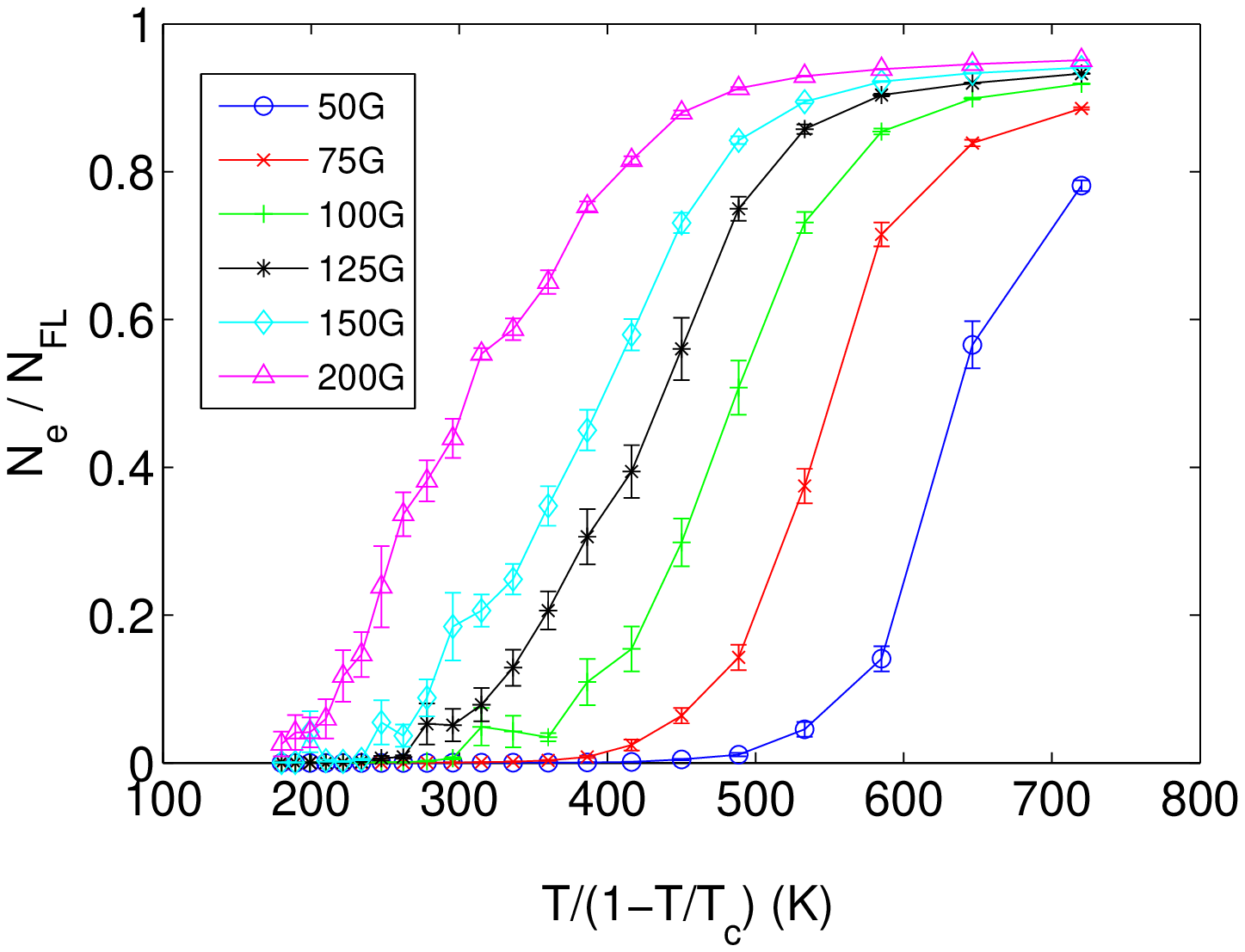}
    \end{minipage}}
\subfigure[]{
    \label{fig01_b}
    \begin{minipage}[b]{0.5\textwidth}
    \centering
    \includegraphics[width=2.8in]{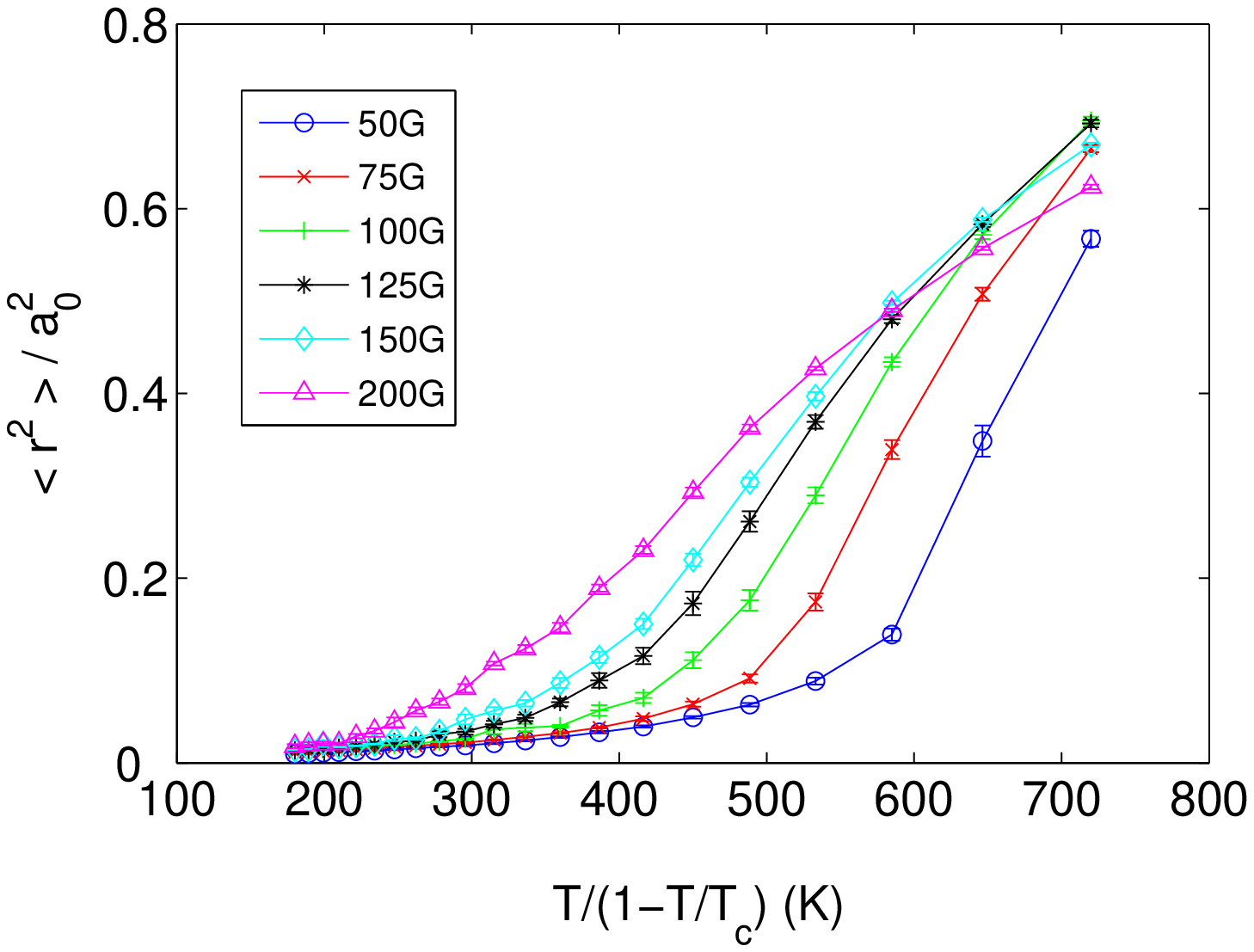}
    \end{minipage}}
\caption{
(color online) Static simulations for systems of 36 layers with $B_{\phi}$ = 50G
and different values of B. The curves are average over 10 CD realizations.
\subref{fig01_a} Fraction of FLs belonging to non-simple loops.
\subref{fig01_b} Mean square deviations of FLs.
}
\label{fig01}
\end{figure}

According to the periodic boundary condition along $z$ direction, the top of each FL
is connected via the Josephson interaction with the nearest pancake in the bottom plane.
Below the melting transition, the pancake usually belongs to the same FL as the top pancake.
We say, the FL form a simple loop. As temperature increases, the thermal motion of the pancakes
starts to make some pairs of FLs get close enough that flux cutting and reconnection
happens, and some FLs become part of a non-simple loop.
Figure \ref{fig01_a} shows the fraction of FLs belonging to non-simple loops.
Below the melting transition, none of the FLs entangle with others and the curves lie exactly at zero;
above the delocalization transition, nearly uniform liquid is formed and the curves approach one.
The criteria for melting can be taken to be that the fraction of non-simple loops reaches $\approx$ 0.1 - 0.2.
It shows that the melting temperature decreases as the magnetic field increases.
Figure \ref{fig01_b} shows the average mean square deviation of the pancakes
on each FL relative to their ``center of mass'' \cite{yyg}.
The results are in units of $a_0^2$ for B = 100G.
The melting transition is characterized by the sharp change of slope. For example, for B = 50G,
the change happens around effective temperature 600, which is about 78K,
with mean square deviation $\approx$ 0.1 - 0.2.

\begin{figure}
\subfigure[]{
    \begin{minipage}[b]{0.5\textwidth}
    \centering
    \includegraphics[width=2.8in]{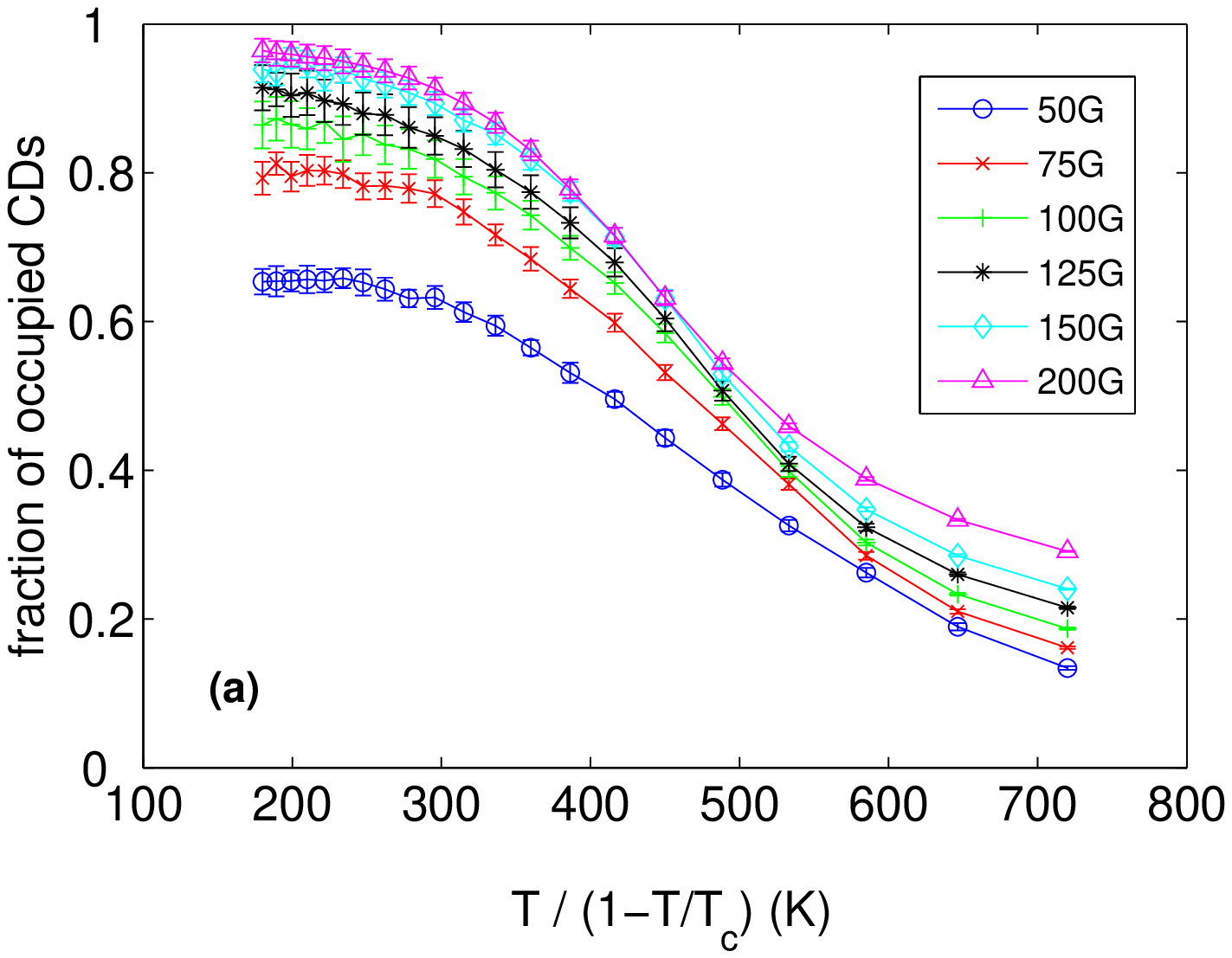}
    \end{minipage}}
\subfigure[]{
    \begin{minipage}[b]{0.5\textwidth}
    \centering
    \includegraphics[width=2.8in]{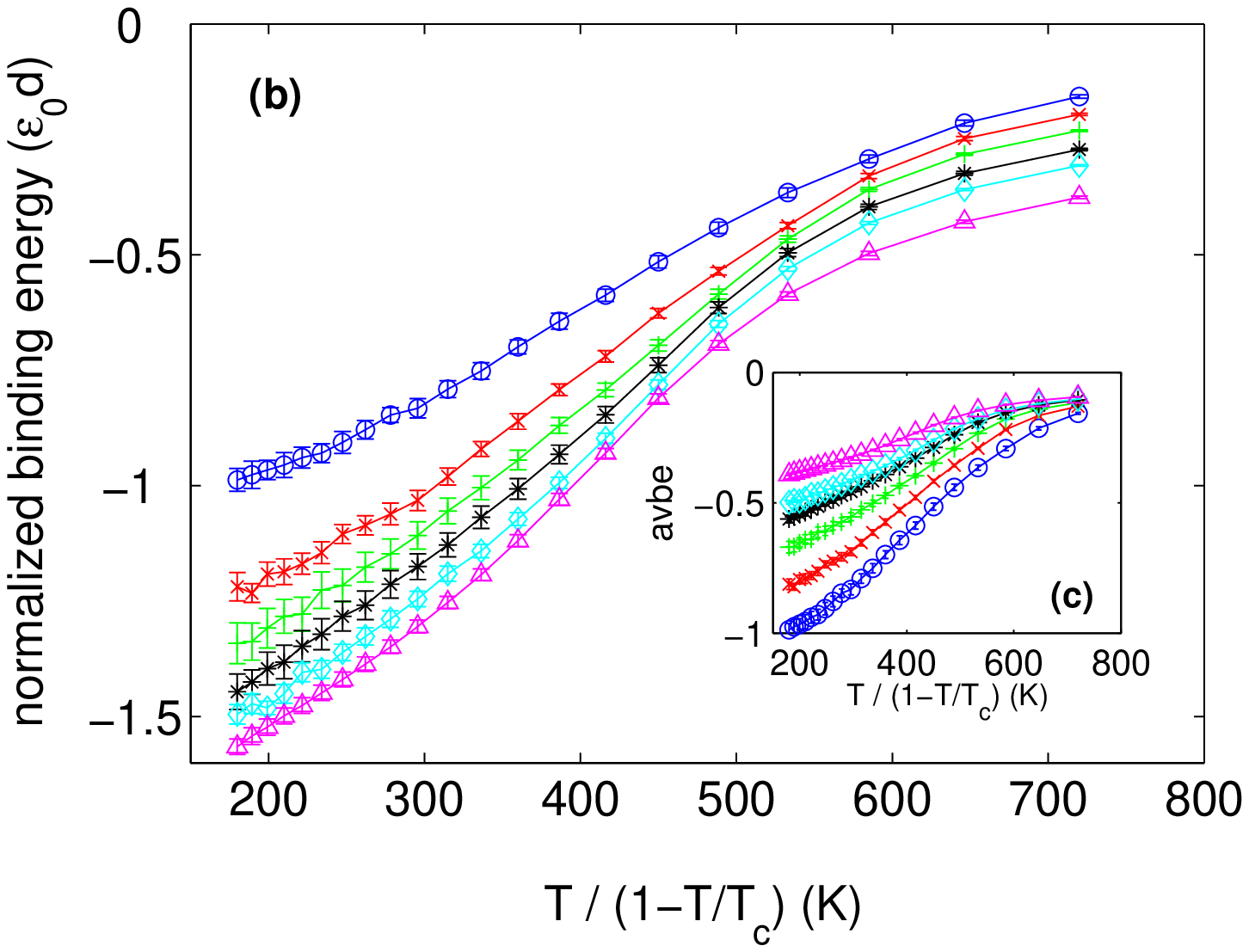}
    \end{minipage}}
\caption{
(color online) Static simulations for systems of 36 layers with $B_{\phi}$ = 50G
and different values of B. The curves are average over 10 CD realizations.
(a) Fraction of occupied CDs.
(b) Normalized binding energy.
(c) Average binding energy per pancake.
(b) and (c) used the colors as in (a) to represent different magnetic fields.
}
\label{fig02}
\end{figure}

Another quantity of interest is the number of trapped pancakes. Usually, each CD
is occupied by a single FL. The trapped FLs form a cage for the interstitial FLs.
Below the melting, the interstitial FLs form crystallites in the pores.
At temperatures slightly above the melting transition, the crystallites
melt locally and form nanoliquid in the pores. Higher temperature enables the trapped
FLs to escape out of the CDs and the cage also melt, so a homogeneous liquid is formed.
To compare results of different magnetic fields,
Fig. \ref{fig02}(a) plots the fraction of occupied CDs, which is given by the fraction
of trapped pancakes multiplied by FL to CD ratio.
As a common character for all the field values, the fractional values decrease as
temperature is increased. The curve for $B$ = 50G is significantly lower than
that of higher field values at low temperatures. Even though the FL to CD ratio is one,
the repulsion between the FLs tries to make them evenly spaced
and some FLs have to be interstitial and some defects become unoccupied.
Figure \ref{fig02}(b) shows the normalized binding energy, which is the binding energy
per plane per CD. As a consequence of Fig. \ref{fig02}(a), the values increase as the temperature
increases because the fraction of occupied CDs decrease.
As a comparison with our previous work \cite{goldcuan},
Fig. \ref{fig02}(c) is the average binding energy per pancake.

\begin{figure}
\subfigure[]{
    \begin{minipage}[b]{0.5\textwidth}
    \centering
    \includegraphics[width=2.8in]{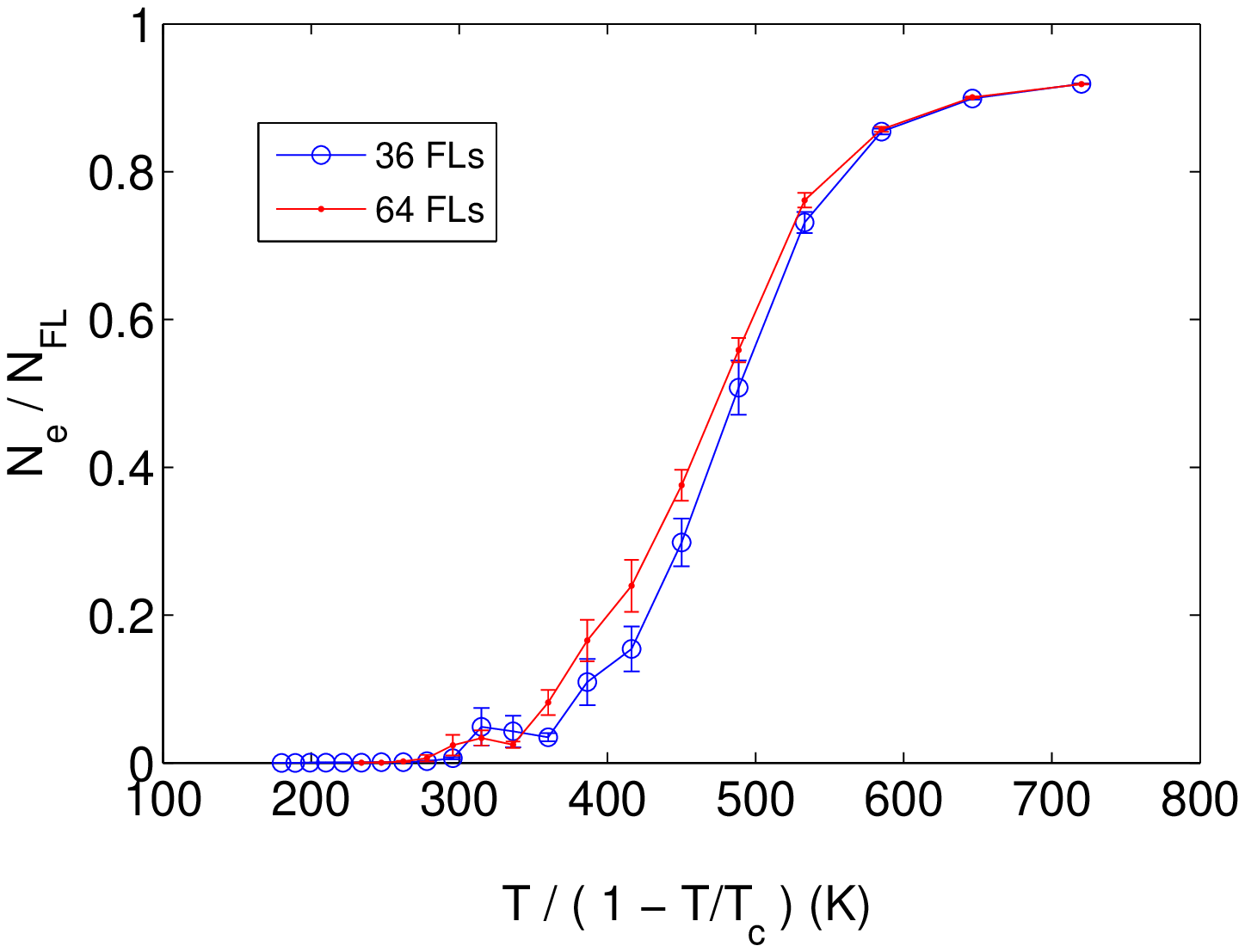}
    \end{minipage}}
\subfigure[]{
    \begin{minipage}[b]{0.5\textwidth}
    \centering
    \includegraphics[width=2.8in]{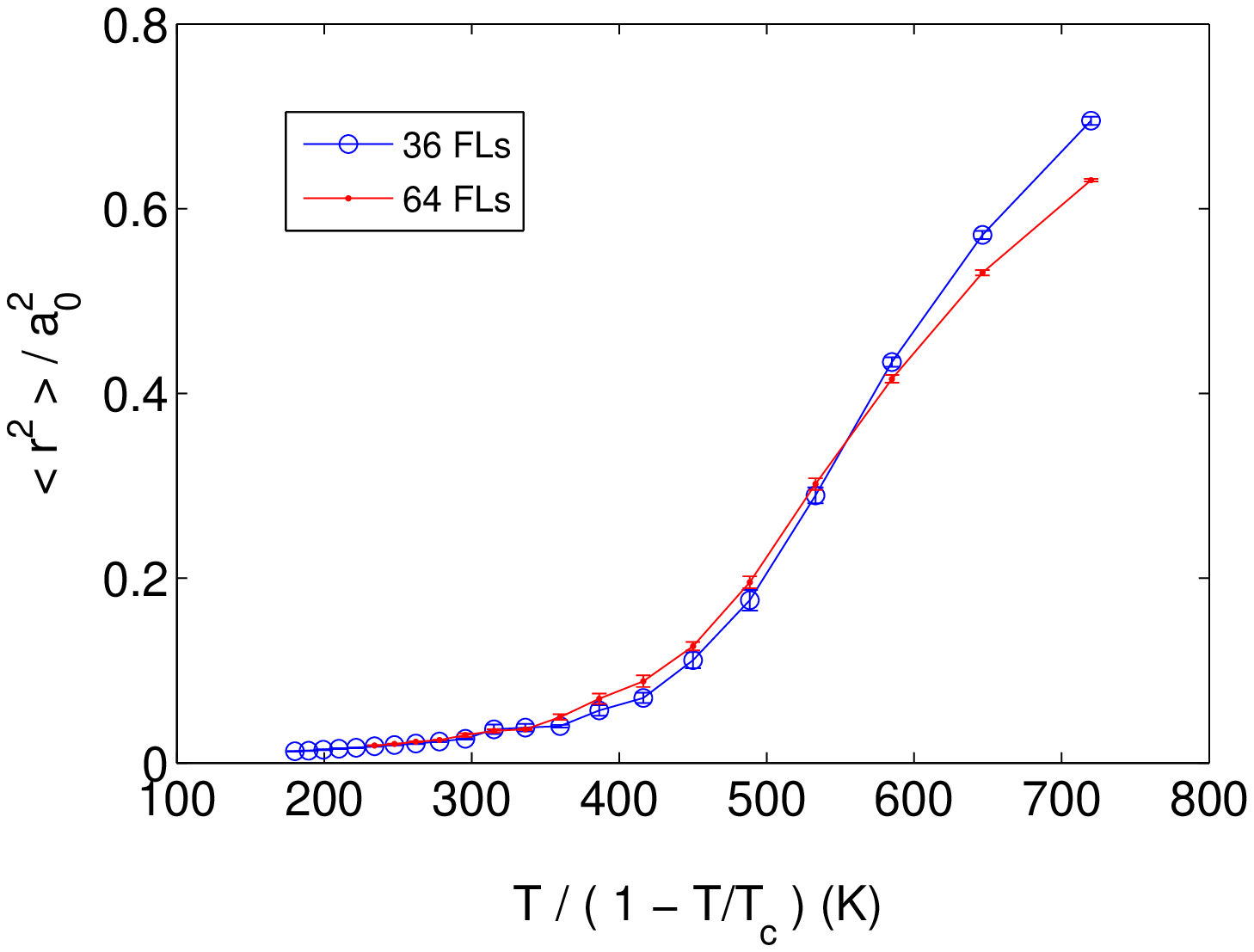}
    \end{minipage}}
\caption{
(color online) Static simulations for systems with 36 FLs (blue curves with circles)
and 64 FLs (red curves). The curves are average over 10 CD realizations.
B = 100G and $B_{\phi}$ = 50G. (a) Fraction of non-simple loop FLs. (b) Mean
square deviations of FLs.
}
\label{fig03}
\end{figure}

In order to verify that our system is large enough and finite size
effects will not significantly change the results, we carried out
simulations with 64 FLs in 36 planes. A comparison between the results
of 36 and 64 FLs is depicted in Fig. \ref{fig03}. We see that the
results are very similar, the transition for 64 lines is slightly
sharper but this is not a significant effect.

Next, we discuss the dynamical measurements which were performed in order to locate
the delocalization transition and the irreversibility line.
The system is also started from a hexagonal lattice of straight and vertical FLs.
After 20 to 50 units of model time for equilibration without driving force,
a uniform in-plane driving force $f_L$ is applied on all the pancakes along the $x$ direction.
We used $f_L =$ 0.025 and 0.1, which are small compared with the binding force on the trapped pancakes.
Five to thirty units of model time were used on equilibration with driving force and 20 to
50 units of model time were used on measurements.

\begin{figure}
\centering
\includegraphics[width=.4\textwidth]{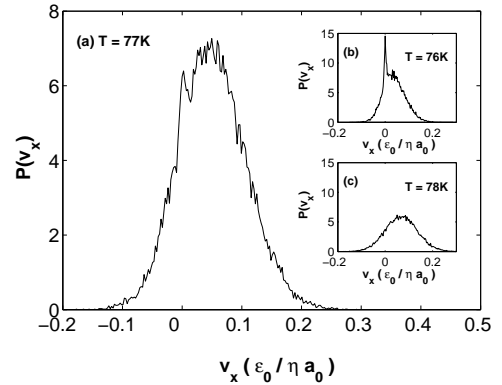}
\caption{
Normalized distribution of average velocity along $x$ direction.
B = 100G, $B_{\phi}$ = 50G, simulation of system with 36 FLs and 200 layers.
The driving force $f_L$ was taken to be 0.1 and 4 different CD configurations were done.
(a), (b) and (c) are the distributions at temperatures in the vicinity of the
delocalization transition.
}
\label{fig04}
\end{figure}

Under a driving force, the interstitial pancakes are much easier to drift relative to the
trapped ones. For driving force $f_L \lesssim 0.1$ and below the delocalization transition,
most of the trapped pancakes remain trapped during the time span of our simulation.
The average velocity of each pancake was measured.
Figure \ref{fig04} shows its distribution for temperatures near the delocalization transition.
The distribution is normalized so that the area of the region under the curve is one.
Below the transition temperature, the distribution is characterized by a sharp peak at zero
and a wider and lower peak centered above zero.
The sharp peak corresponds to the trapped pancakes
and the wide peak corresponds to the interstitial ones. The driving force we used was small
enough that the trapped pancakes would not be pulled out of the columnar defects below
the delocalization transition but large enough that the interstitial pancakes
could move through the skeleton formed by trapped FLs. As temperature increases,
the height of the sharp peak decreases and the wide peak becomes wider.
The sharp peak disappears above the delocalization temperature and the center of the remaining
Gaussian shaped wide peak approaches 0.1, which is the value of the driving force.

One should notice that the width and height of the distribution in Fig. \ref{fig04}
depends on the length of the measurement. The longer the measurement is, the narrower and taller the wide peak is.
For random walk, the average distance the walkers travel is proportional to the square root
of time, thus the average velocity of the walkers decreases as the time of averaging becomes
longer. The movement of the interstitial pancakes are different from a simple random walk.
They are confined by other FLs, especially by the trapped ones.
They are also confined by pancakes in adjacent layers through Josephson interaction.
However, our simulations do show that the width of the wide peak
times the square root of the time span of the measurement is roughly constant
near and above the delocalization transition.

\begin{figure}
\centering
\includegraphics[width=.4\textwidth]{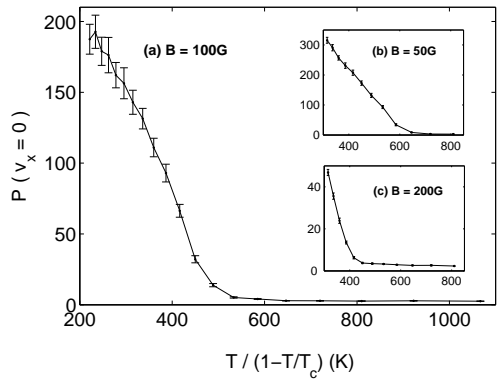}
\caption{
Probability density of zero average velocity at various temperatures.
Results for three different fields are shown here.
Simulations on systems with 36 FLs and 36 layers. $B_{\phi}$ = 50G, driving force $f_L$ = 0.1.
Average over 10 different CD configurations.
}
\label{fig05}
\end{figure}

In Fig. \ref{fig05}, the probability density of zero average velocity,
{\it i.e.} the value of the average velocity distribution at zero, is plotted.
Results for magnetic fields of 50G, 100G and 200G are shown.
The delocalization transition is characterized by the abrupt
change of the slop of the plotted curves, which happens around effective temperatures
600, 500 and 400 respectively, corresponding to 78K, 76K and 73K.
The delocalization transition temperature decreases as the magnetic field increases.

\begin{figure}
\centering
\includegraphics[width=.4\textwidth]{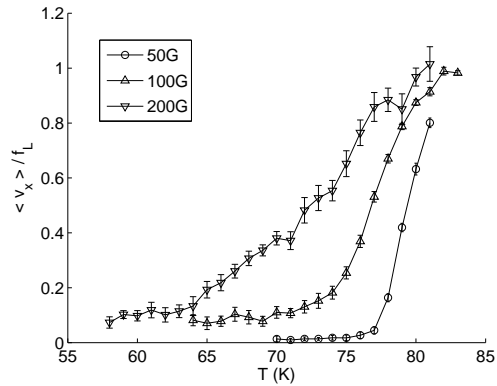}
\caption{
Average drift velocity over driving force. Results with
magnetic fields of 50G, 100G and 200G are shown.
Used the same data as in Fig. \ref{fig05} for B = 50 G and B = 100 G, where, $f_L = 0.1$.
For B = 200 G, $f_L = 0.025$ was used.
}
\label{fig06}
\end{figure}

It is possible to locate the irreversibility line from the time averaged drift velocities,
by taking an average over all pancakes. The average velocity of the pancakes is depicted
in Fig.\ref{fig06}. For temperatures below the irreversibility line, which is the
threshold of mobility as a function of temperature, the curves are flat and the
drift velocity is small. Then at a temperature about equal to the melting temperature,
their slopes increase sharply. The sharp rise in mobility is most clear for the
$B = 50$G case, for which the melting transition and the delocalization transition
happen at the same temperature. At low temperatures, the FLs are pinned collectively
by the CDs. They are nearly immobile under a small driving force. When the temperature
is raised high enough, that the interstitial pancakes can move easily in the pores,
a small driving force will make the FLs drift, signaling the onset of depinning.
Above the delocalization temperature, the trapped pancakes also start to drift under
the small driving force, making the ratio of average drift velocity over driving
force approach one.

\begin{figure}[h]
\centering
\includegraphics[width=.45\textwidth]{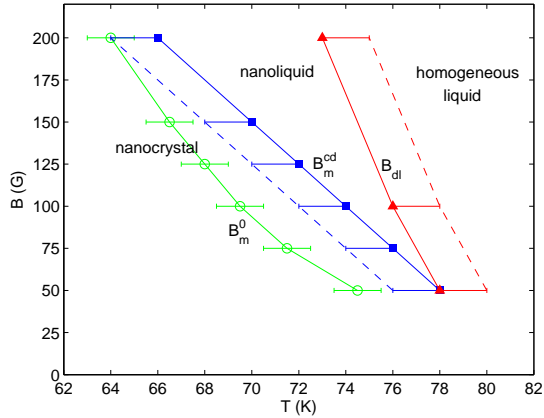}
\caption{
(color online) Phase diagram in T-B plane. The melting transition $B_m^0$ of pristine system,
the melting transition $B_m^{cd}$ of system with columnar defects ($B_{\phi}$ = 50 G) and
the delocalization transition $B_{dl}$ of system with columnar defects ($B_{\phi}$ = 50 G) are shown.
}
\label{fig07}
\end{figure}

The results above are summarized into a phase diagram (Fig.\ref{fig07}). It is divided into
three regions by the melting line $B_m^{cd}$ and the delocalization line $B_{dl}$.
Between these two lines, the vortex system is in a nanoliquid phase, in which the interstitial
FLs are entangled and melted but are separated from the trapped FLs. To the left of the dashed line
associated with the melting line, the system is in the nanocrystaline phase, in which all the FLs are separated
from each other, the interstitial ones form triangular lattices in the pores
surrounded by the trapped ones. The error bars indicate the width of the transition
region due to the finite size of our system. The melting line $B_m^0$ of a pristine
system is also plotted for comparison. To the right of the dashed line associated with the delocalization line,
the system is in a homogeneous liquid phase, where the effect of the CDs becomes less important
and all the FLs become entangled.

\section{tilted columnar defects}

It is interesting to study the effect of the tilting of CDs on the properties
of the vortex system. Hwa, Nelson and Vinokur \cite{hwa}, Nelson and
Vinokur \cite{Nelson,NV}, and more recently Refael et al. \cite{refael} studied the
effect of a tilted magnetic field on the properties of the Bose glass
system. In their investigation the CDs are along the c-axis and the
magnetic field has a component $H_\perp$ along the a-b planes, thus
forming an angle $\theta$ between the direction of the applied field
and the CDs. When the angle $\theta$ is small the vortices which are localized
on the CDs are completely trapped along the CDs and the Bose glass
phase is preserved. At larger $\theta>\theta_c$ ($H_\perp>H_\perp^{c}$), the
vortices form a staircase structure hopping from one CD to the
next. For different vortices those kinks tend to align in ``chains''.
For an even larger angle $\theta>\theta_a$ the vortices follow the
field direction and are essentially unaffected by the correlated
nature of the CDs. This picture appears to describe correctly the
situation in YBCO in the Bose glass regime ($B<B_{\phi}$) and is supported
by various experiments, both for thermodynamical \cite{hayani} 
and dynamical properties \cite{civale,silhanek}.

For BSCCO which is much more anisotropic the situation appears more
complicated and less clear \cite{drost,kameda}. Furthermore in this case there are
interesting results on the nanosolid phase which exists for $B>B_{\phi}$
and is different from the Bose glass phase. There are no theoretical
investigations on how the nanosolid phase and its two stage melting
transition might be affected by a tilting angle between the direction
of the CDs and the magnetic field. To try to answer this question we
recently carried out experiments and numerical simulations to try to
understand the properties of such a system \cite{nurit}. Here we give
more details on the method and the results of the simulations.

In this work we limited our investigation to the case of a vertical external
magnetic field (parallel to the $c$ axis), and considered only the
case of tilted CDs at various angles. Simulations with 36 layers are not enough any more,
since the shift of the top of the CDs relative to their bottom would be too small.
We used 200 layers, which is still manageable by the current computing power.
Most of the simulations were done on systems with 36 FLs and 18 CDs.
The magnetic field was set to be 100G. For comparison, both systems with vertical CDs
and systems with tilted CDs were simulated.
The comparison between the two systems was done for the same value of
$B^{\rm eff}_{\phi}=B_{\phi}\cos\theta$, in order to match the
experimental situation in which the ratio of vortices to CDs remains
fixed as $\theta$ changes \cite{nurit}.
Both static and dynamical measurements were carried out and the results
are averaged over four different random CD configurations. 

\begin{figure}
\centering
\includegraphics[width=.45\textwidth]{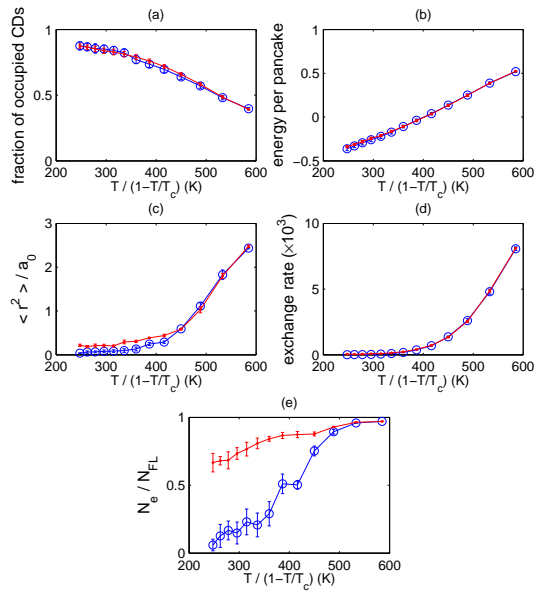}
\caption{
(color online) Static measurements on systems with vertical CDs (blue curves with circles) and titled CDs
(red curves, CDs are tilted by $45^\circ$ away from $c$ axis).
(a) Fraction of occupied CDs.
(b) Total energy per pancake.
(c) Mean square deviation of FLs.
(d) Exchange rate.
(e) Fraction of FLs belonging to non-simple loops.
}
\label{fig08}
\end{figure}

Figure \ref{fig08} shows the results of static measurements.
Figure \ref{fig08}(a) shows the fraction of occupied CDs,
{\it i.e.} the number of trapped pancakes over the number of trapping sites.
Over the temperature range of our simulations, the vertical case and the tilted case give the same results within
the error bars. At effective temperatures below 350, the ratio is close to one, so the CDs are almost fully occupied.
In Fig. \ref{fig08}(b), the total energy per pancake are almost the same in the
two cases. In Fig. \ref{fig08}(c), the mean square deviation of the FLs in the tilted case
is a little higher than that of the vertical case at low temperatures. Since our measurements are
relative to the center of mass of each FL, and as will be shown later, the FLs are tilted by the tilted CDs,
the deviations are raised to higher values.
The sudden change in the curves' slopes reflects the position of the melting transition
of the interstitial FLs. Figure \ref{fig08}(c) might suggest a one degree increase in the
melting temperature for tilted CDs, but this is within the simulation error.
Figure \ref{fig08}(d) shows the number of ``exchanges'' per unit
time, i.e. the rate of cutting and recombining of FLs. This number
is under 1000 below the melting transition and increases rapidly above
the melting. We see that it is the same for both straight and tilted CDs.
Figure \ref{fig08}(e) shows the amount of entanglement, {\it i.e.} the fraction
of non-simple loops. The flat area of the curves is an indication for the melting
transition. First, we note that for straight CDs, the fraction of non-simple loops at
the transition for 200 layers is higher than the corresponding value for 36 layers,
which is about 0.2. This is because for longer FLs, there is more chance to cross and also
to form occasional kinks involving the CDs. This figure is the only one that shows
a significant difference between the straight and tilted CDs in terms of the absolute
value of their entanglement. This is likely due to the fact that FLs
trapped on CDs are tilted and so are those FLs in the vicinity of a CD.
Thus the top and bottom of those FLs are not adjacent and they may
connect to other FLs to form loops. This interpretation is favored by the fact that
the number of exchanges for the tilted CD case is not larger than for
the case of straight CDs but the amount of their entanglement is
larger nevertheless. Thus this criterion (of entanglement) using the
boundary conditions may not be a good indication for the melting
transition in the case of tilted CDs 
as it is for the case of pure system or a system with straight CDs. The flat plateau
region of the curve should give a rough indication where the melting takes place.
Thus for tilted CDs, we base our criterion for the melting on the transverse
fluctuations of the FLs and on the number of exchanges per unit time
rather than on the amount of entanglement.

We now account for the energies involved: The electromagnetic energy cost per layer
associated with the tilting of an infinitely long stack of pancakes is given by \cite{clem}
$\epsilon_0 d \ln((1+\cos{\theta})/2\cos{\theta})$. This cost is exceeded,
for angles less than about $75^{\circ}$, by the energy gain from the pinning of
columnar defects, which is of the order of $\epsilon_0 d$ per layer. In addition,
one should also include the energy $-\mathbf{B}\cdot \mathbf{H}/(4\pi)$, which
favors to align the average vortex orientation with $H$,
which is implemented in the simulation by
the periodic boundary conditions: When  
a stack is tilted, the stacks of images along the $z$ direction are not positioned
along the tilted direction but are above (and below) the center of mass of the 
tilted stack and because of the electromagnetic interaction they pull
the stack in the z-direction, acting like the $-\mathbf{B}\cdot
\mathbf{H}/(4\pi)$ term. The competition between the binding energy
gain and the loss of electromagnetic energy associated with a tilt
favors the formation of kinks. The trapped FLs tilt to take advantage
of the binding energy of the tilted CDs, and the interstitial FLs tilt
because of repulsion by the trapped pancakes. But in order to relieve the
frustration of the electromagnetic interaction, kinks are formed and the FLs jump
back to align with the z axis which is the external field
direction. The larger the tilting angle of the CDs the more kinks
appear along the FLs. The kinks on neighboring FLs tend to align with
each other.
The formation of the kinks increases the Josephson interaction part,
however, in high anisotropic materials like BSCCO, this cost is negligible compared
with the total energy \cite{bulaevskii3}.
Near the transition the
transverse fluctuations due to the tilt-kink combination is not much
different from the transverse fluctuations due to temperature fluctuations.
Our simulations show that the vortices form a tilted rigid matrix which is
parallel to the CDs, and the CDs are almost fully decorated by pancakes.
Because of the kinks, the interstitial vortices can switch cage after a certain
distance along the $z$ direction, but each section is still caged by the trapped
pancakes on the CDs. Also interstitial FLs can themselves by trapped on CDs for
some portion of their length. In the vicinity of the melting transition ($T$ =
73K to 77K), considering the error bars of the measurements, we found that
systems with CDs tilted at $45^{\circ}$ have nearly the same total energy as those with
non-tilted CDs. One can conclude that the enhanced caging potential created by
the CDs, which raises the melting temperature \cite{banerjee}, is not
significantly affected when increasing the angle between $\mathbf{B}$ and the
CDs, thus the thermodynamic properties, {\it i.e.}, the melting transition and
the delocalization, are angle independent.
This has been shown in our static simulations in Fig. \ref{fig08}, and will be
further demonstrated for the delocalization transition in the following discussion.

\begin{figure}
\centering
\includegraphics[width=.4\textwidth]{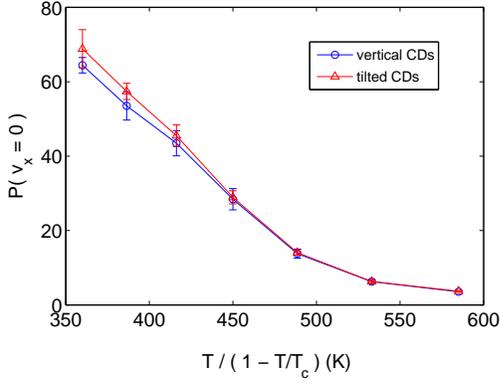}
\caption{
(color online) Probability density of zero average velocity. the blue curve with circles is for vertical CDs and
the red curve with triangles is for tilted CDs. Driving force $f_L = 0.1$.
}
\label{fig09}
\end{figure}

The effect of tilting of the CDs on the delocalization transition is also investigated.
Figure \ref{fig09} shows the probability density of zero average velocity for both
vertical case and titled case. The two curves coincide within
the range of the error bars, especially at temperatures near the delocalization transition.
The average velocity distributions of these two cases are also compared and one
cannot see any significant difference. One can conclude that the tilting of CDs does not have noticeable
effect on the delocalization transition of the vortex system.

\begin{figure}
\centering
\includegraphics[width=.5\textwidth]{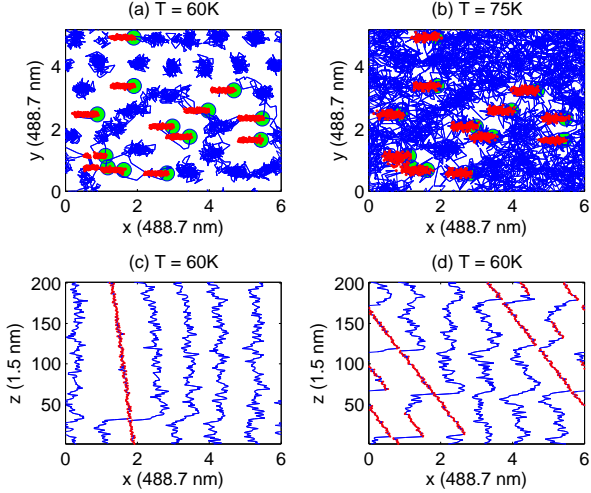}
\caption{
(color online) Snapshots of vortex stacks and CDs tilted at $45^{\circ}$ (a, b and c)
and $80^{\circ}$, $B_{\phi}^{\rm eff} = 36$ G. (d) for $B$ = 100 G. Pancakes belonging to the same stack are connected.
Free pancakes are in blue and trapped ones in red. Only CDs at the bottom layer are shown (green).
(a)(b) Projection onto a-b plane (top-view). (a) Nanosolid phase.
(b) Nanoliquid phase. (c) Projection of (a) onto a-c plane (side view), first row is shown.
(d) Side view with CDs at $80^{\circ}$.
}
\label{fig10}
\end{figure}

Snapshots are shown here on systems without a driving force.
Figure \ref{fig10}(a) and \ref{fig10}(b) shows the top view of a system with CDs tilted
by $45^{\circ}$ at 60K and 75K respectively. At 60K, the system is in the nanosolid
phase and individual FLs can be distinguished clearly. The interstitial FLs
try to form hexagonal lattices. At 75K, the system is in nanoliquid phase.
Interstitial FLs become entangled and the pancake vortices distribute nearly
uniformly in the pores. But the temperature is not high enough to make trapped
pancakes delocalize yet. The nanoliquid is repelled by the trapped FLs, thus forming
void spaces surrounding the CDs.
Figure \ref{fig10}(c) is a side view of the first row of the FLs in Fig. \ref{fig10}(a).
Usually, a CD is occupied by more than one FL and kinked structures emerge.
The kinks go along the a-b plane and extend for less than ten CuO$_2$ layers. There are also kinks
on interstitial FLs. They tend to reside at the same layers with those attached
to CDs and form trains of kinks \cite{hwa}, which resemble the Josephson and Abrikosov
crossing lattices \cite{koshelev2} found in BSCCO with field tilted with respect to the $c$ axis.
Another feature is that the interstitial FLs break into segments, which are
titled to some degree along the direction of the CDs.
Because of the kinks, the FL segments jump back periodically, keeping the FLs
vertical on average. At higher temperatures,
The kinked structures and tilting of interstitial FLs still exist but are complicated by the entanglement of the FLs.
In Fig. \ref{fig10}(d), the CDs are tilted by a larger angle than in Fig. \ref{fig10}(c),
as a consequence, more kinks appear and the separation between the trains of kinks becomes smaller.

\begin{figure}
\centering
\includegraphics[width=.4\textwidth]{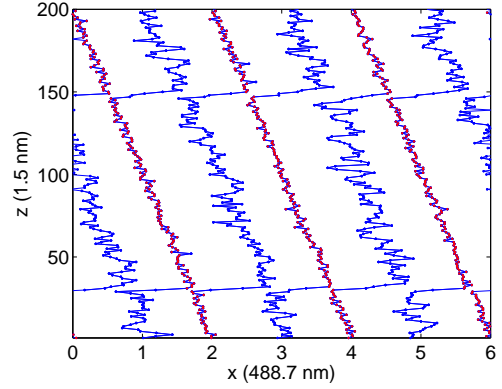}
\caption{
(color online) Snapshot of a 2D system with 200 planes, 6 FLs and 3 CDs.
Pancakes belonging to the same stack are connected. Trapped ones are in red and interstitial ones are in blue.
CDs are evenly spaced and titled at $73^{\circ}$ so that they are continuous at the boundaries. T = 70 K.
}
\label{fig11}
\end{figure}

Under large enough driving force and temperature in the proper range,
FL kinks can move along $z$ direction and the pancake vortices drift along the direction of the driving force.
To show this, we simulated a two-dimensional system with six FLs and three titled CDs (Fig. \ref{fig11}).
The pancakes were confined to move only in the $x$-$z$ plane. 
The CDs were arranged such that they are continues at the boundaries.
We initialized the FLs to be vertical, straight and evenly spaced.
After simulating for enough time to the let the system equilibrate
without a driving force, two trains of kinks were formed.
A driving force of magnitude $f_L$ = 0.5 was then applied along the positive $x$ direction. 
At 70 K, the kinks moved downward altogether with a constant speed of about 13 CuO$_2$ planes
for every ten time units.

\section{Conclusions}
To conclude we have used Langevin molecular dynamics simulations to
investigate the melting transition of the vortex matter in the
presence of vertical CDs and in the presence of tilted CDs. For
vertical CDs our results are in full agreement with our previous
multilevel MC simulations reported in Ref.[\onlinecite{goldcuan}]. 
Furthermore we measured also dynamical
properties of the system under an external current that applies force
on the vortices and these located the melting and delocalization
transitions at the same temperatures as the static measurements,
giving a more precise identification of the location of the
delocalization transition. For the melting transition the dynamical
measurement actually identifies the irreversibility line associated
with partial depinning, and this occurs at the same temperature as the melting
for the case of vertical CDs.

For the case of tilted CDs, the porous vortex matter in BSCCO is shown
to preserve its thermodynamic properties even when the field is tilted
away from the CDs. The positions of the  melting and delocalization
lines remain unchanged while the tilting angle of the CDs vary, at
least up to angles of $45^\circ$. Our simulations show that increasing the
angle between the field and the CDs leads to formation of weakly
pinned vortex kinks, while preserving the basic structure of a rigid
matrix of pancakes residing along the CDs with nanocrystals 
of interstitial vortices embedded within the pores of the matrix.
We also showed that under the influence of an applied force (like a
current) the kinks slide down and lead to a global motion of the
pancakes in the direction of the applied force. Thus we expect that
for large tilting angle between the CDs and the magnetic field, which
result in the proliferation of kinks, the irreversibility line should
move to lower temperature, closer to the location of the melting in
the pure system (no CDs). To measure this shift explicitly in the
simulations as a function of the angle requires much more
computational time than currently available, but it was verified in the
recent experiments \cite{nurit}.  

\section{Acknowledgments}
This work was supported by the US
Department of Energy (DOE) under grant No. DE-FG02-98ER45686. We
also thank the DOE NERSC program and the Pittsburgh 
Supercomputing Center for time allocations. YYG thanks the Weston
Visiting Professors program of the Weizmann Institute of Science for support
during a recent visit and E. Zeldov for his hospitality.
YYG also thanks N. Avraham and E. Zeldov for useful discussions.


\end{document}